\documentstyle[aps,preprint,prc,epsf]{revtex}

\newcommand{\be}{\begin{equation}}
\newcommand{\ee}{\end{equation}}
\newcommand{\ba}{\begin{eqnarray}}
\newcommand{\ea}{\end{eqnarray}}
\newcommand{\bb}{\begin{thebibliography}}
\newcommand{\eb}{\end{thebibliography}}
\newcommand{\bi}[1]{\bibitem{#1}}

\begin{document}
\draft
\title{ Can Nuclear Decay constant be Modified ?  }
\author{ Il-Tong Cheon}
\address{ Department of Physics, Yonsei University, Seoul 120 -- 749,
Korea }
\maketitle
\begin{abstract}
The life-time of $^{133}Cs$ in the first excited state has been measured
with the M\"ossbauer method by placing the absorber, $CsCl$, between two
parallel flat plates and the gamma-ray source, $Ba^*TiO_{3}$, in the free space.
The result is 9.37$\pm$0.19 ns, which is 49 percent larger than the standard value 6.27$\pm$0.02 ns.
\end{abstract}
\vskip 1cm
\pacs{PACS numbers :  21.10. Tg, 23.20. Lv, 25.20. Dc, 27.60.+j}

\section{Introduction}

 Recently, a series of investigations were carried out on the effect of
 vacuum fluctuation on nuclear energy levels[1-4]. These works explored
 energy level shifts due to vacuum fluctuations in a finite space.
 In addition, it was reported that the life-time of the hydrogen
 $2P$-state could change by about 3$\sim$5 percent, if the atom was
 placed between two parallel conducting plates separated from each
 other by $1\mu m$ [5]. There are another reports that spontaneous
 emissions by a Rydberg atom [6] and those of cyclotron radiation [7]
 can be inhibited by cavity effects. There are also some attempts
 to observe changes of decay constants with technetium (see Table
 1).

 This letter will report an experimental result that the life-time of
 a radioactive nucleus put between two parallel flat plates becomes
 longer than that of the nucleus in  free space. To date, the life-time
 of a nucleus has been believed to be an invariable quantity. Namely,
 it has been supposed that even if electric or magnetic field were
 applied, nuclear energy level widths would remain unchanged, although
 shifts or splitting could take place. Notice that the life-time is
 proportional to the inverse of the energy level width.

\section{Experiment}

 In order to explore the shift of nuclear life-time, we carried out
 M\"ossbauer measurement of the width of the first excited state in
 the $^{133}Cs$ nucleus using the facilities at Leuven University.

\subsection{Gamma-ray source}
 The compound $Ba^*TiO_{3}$ was manufactured as the gamma-ray source.
 The manufacturing process is as follows : A mixture of $Ba^*Cl_{2}$,
 $BaCO_{3}$ and $TiO_{2}$ in the ratio of 1:3:4 was pulverized and
 heated at 1200$^{o}$C for 20 hours in an electric oven followed by
 annealing at 700$\sim$800$^{o}$C for two days.
 Since it was not sufficiently hard at this stage, we pulverized it
 again and, then, heated once more at 1200$^{o}$C for 40 hours.
 X-ray diffraction analysis showed a characteristic $BaTiO_{3}$
 compound pattern and confirmed this sample to be a good gamma-ray
 source.

\subsection{Measurement without plates}
 The source in 2.4mCi amount was fixed to the electromechanical
 spectrometer providing the Doppler velocity. The result obtained with
 the scintillation counter showed indeed a typical M\"ossbauer
 spectrum of a single line.
 This spectrum was obtained at 4.2$\rm K$ by using $CsCl$ powder
 with 300$mg/cm^{2}$ thickness as the absorber.
 Its volume density was 4$g/cm^{3}$, namely 300$mg$ $CsCl$ in a
 volume $1cm\times1cm\times0.75mm$.
 The line broadening is 0.71$mm$/$s$ and the relative depth of the
 spectrum is around 1.3 percent.
 The spectrum corresponds
 to an 81KeV gamma-ray emitted by the transition from the first excited
 state $\frac{5}{2}^{+}$ to the ground state $\frac{7}{2}^{+}$ of
  $^{133}Cs$. The reduced $\chi$-square of the  Lorentzian spectrum
 between 6 and 251 channels was 1.3, and the relative error for line
 position is around $2\%$. The width at the half-height of the
 M\"ossbauer spectrum was $\Gamma_{exp} = 0.796\pm0.014mm/s$, equivalently
 $\Gamma_{exp}=(2.149\pm0.038)\times10^{-7}$eV.
 This value contains a thickness effect[8] of the absorber and
 possible unresolved hyperfine interaction arising from distortions in
 the $BaTiO_{3}$ lattice and a possible experimental broadening effect
 due to vibration. If the source is free from the thickness
 effect, the  width of the M\"ossbauer spectrum is usually given
 with the natural line width, $\Gamma_{nat}$, as
\ba
 \Gamma_{exp}=(2+0.270t_{a})\Gamma_{nat} + \Delta\Gamma,
\ea
 where $\Delta\Gamma$ is the systematic broadening and the
 thickness effect of the absorber is expressed by a factor
 $0.270t_{a}$. Here $t_{a}=n\sigma_{0}f$ with n being the number of
 radioactive nucleus per unit area, $\sigma_{0}$ the maximal cross
 section for resonant nuclear absorption and $f$ the recoilless
 fraction of the absorber. For our absorber $CsCl$, we have
 $n=1.075\times10^{21}cm^{-2}$, $\sigma_{0}=1.021\times10^{-19}cm^{2}$
 and $f=0.0145$[9]. The source $BaTiO_{3}$ might not be free from the
 thickness effect. For such a case, Eq.(1) should be converted into a
 form
\ba
 \Gamma_{exp}=\xi_{s}\Gamma_{nat}+(1+0.270t_{a})\Gamma_{nat} +
 \Delta\Gamma,
\ea
where $\xi_{s}=1+0.270\tilde{t}_{a}$ with
$\tilde{t}_{a}=\tilde{n}\tilde{\sigma}_{0}\tilde{f}$. The value of
$\xi_{s}$ is not known at this moment, but it does not matter for the
investigation of the effect of plates because the term depending
on $\xi_{s}$ disappears in the final expression. This fact can be
seen later.

\subsection{Measurement with plates}
 Let us now explore the case of the absorber placed between
 two parallel flat plates. We
 prepared the plates in the following manner.
 Silicon wafer plates of 3$\times$3$cm^{2}$ with 0.58$mm$ thickness
 were coated by gold with about 100${\AA}$ thickness at room temperature
 using an evaporation method developed by the Surface Physics Group at
 Yonsei University. The roughness of the plate surface was on the order
 of 0.01$\mu m$.
 The two plates were separated by 0.61$mm$ with stainless stick
 spacers. Accuracy of parallelness was around 2$\mu m$.
 The absorber, $CsCl$(90.0$mg$), was formed in a very thin mylar
 square bag of $1cm\times1cm\times0.15mm$
 at a volume density of $6 g/cm^{3}$, equivalently $90 mg/cm^{2}$
 thickness which yields $n\rightarrow n'=1.613\times10^{21}cm^{-2}$
 and was centered between the two plates to prevent it
 from touching the plate surfaces. Thermal effects
 were very small, i.e., the second-order Doppler shift was negligible,
 since the temperature variation during the experiment was $\pm$0.2K.
 In this experiment, we took such a geometry as the gamma-ray came
 along the direction perpendicular to the flat plates.

 In this experiment, we obtained a very distinct and
 thin spectrum. The width of Lorentzian spectrum is
 $\Gamma'_{exp}=0.707\pm0.014mm/s$, equivalently
 $\Gamma'_{exp}=(1.909\pm0.038)\times10^{-7}$eV.
 In order to extract the effect of plates set around the absorber,
 let us rewrite Eq.(2) in the form
\ba
 \Gamma'_{exp}=\xi_{s}\Gamma_{nat}+(1+0.270t'_{a})\Gamma^{(a)} +
 \Delta\Gamma,
\ea
where $t'_{a}=n'\sigma_{0}f=(1.613\times10^{21}cm^{-2})\times
(1.021\times10^{-19}cm^{2})\times0.0145=2.388$ and $\Gamma^{(a)}$
is the natural line width modified by the plates. Subtracting Eq.(2)
from Eq.(3), we find
\ba
 \Gamma^{(a)}=\frac{\Gamma'_{exp}-\Gamma_{exp}+(1+0.270t_{a})\Gamma_{nat}}{1+0.270t'_{a}}
\ea
 The half-life of the first excited state $\frac{5}{2}^{+}$ in
 $^{133}Cs$ is known to be $\tau_{1/2}=6.27\pm0.02$ ns[10], from which
 the natural line width,
 $\Gamma_{nat}$, can be calculated as
\ba
 \Gamma_{nat}=\frac{\hbar}{\tau_{1/2}}ln2=(0.728\pm0.002)\times10^{-7}\rm eV.
\ea
Substituting $\Gamma'_{exp}=(1.909\pm0.038)\times10^{-7}\rm eV$, $\Gamma_{exp}=(2.149\pm0.038)\times10^{-7}\rm
eV$ and $\Gamma_{nat}=(0.728\pm0.002)\times10^{-7}\rm eV$ into
Eq.(4), we obtain
\ba
 \Gamma^{(a)}=(0.487\pm0.010)\times10^{-7}\rm eV,
\ea
from which the half-life modified by plates can be found as
\ba
 \tau^{(a)}_{1/2}=\frac{\hbar}{\Gamma^{(a)}}ln2=(9.37\pm0.19)\rm ns.
\ea
This value is larger than $\tau_{1/2}=6.27\pm0.02\rm ns$ by
49.4$\%$.

\section{ Discussion }
 Hence, our finding is that the life-time increases by 49.4$\%$ when
 the absorbing nucleus is placed between two parallel
 flat plates. The change of life-time, $\Delta\tau=3.10\pm0.19$ ns,
 arises purely from the effect of plates, because all
 experimental conditions are the same except for setting plates
 around the absorber.

 Why does the life-time become longer when a nucleus is placed between
 two parallel flat plates?
 It may be understood as a phenomenon caused by the
 self-interaction that photons (not necessarily real photons) emitted
 from the excited nuclei are reabsorbed by these nuclei after being
 reflected by plates.

 Consider an excited nucleus put between two parallel flat plates.
 This excited nucleus emits virtual (even real) gamma-rays in arbitrary
 directions, and parts of them are reflected by plates. Since the plate
 is, of course, not perfect, some of the gamma rays are absorbed or
 pass through the plate.
 And the reflected gamma-rays may come back to be
 reabsorbed by the nucleus. Through such a process, the population
 of the excited state in a nucleus could be amplified.

 However, one may worry about that $81KeV$ is too high energy for the
 gamma-ray to be reflected by the plates. Since the wavelength
 of $81KeV$ gamma-ray is about 0.015$nm$, the
 silicon plate would be almost transparent at such a short wavelength.
 Indeed, our Monte-Carlo simulation with the program GEANT3 shows that
 only 0.018 percents of the $81KeV$ gamma-ray are reflected at the
 same energy. Nevertheless, if the process can be repeated many times,
 the effect must be enhanced.

 Generally, the amount of nucleus decaying during $\Delta t$ is given
 by $\Delta N=-\lambda N\Delta t$, from which we obtain
 $ N=N_{0} exp(-\lambda t)$, where $\lambda$
 is the decay constant and $N_{0}$ is the initial value of $N$.
 If emitted photons can once return after being reflected by the
 plates, the equation is modified as
 $\Delta N=-\lambda N \Delta t + \sigma\lambda N\Delta t=-(1-\sigma)\lambda N\Delta t$, where $\sigma$ is
 the reflection coefficient, i.e. $\sigma =0.00018$ for the present
 case. If such a process is assumed to repeat $n$ times, we have

\ba
\Delta N=-(1-\sigma)^{n}\lambda N\Delta t
\ea

For $\sigma =0$, it reduces to $\Delta N =-\lambda N\Delta t$.
Furthermore, $n=0$ leads Eq.(8) to $ \Delta N=-\lambda N \Delta $.
This is the case without any plate, i.e.
photons have no chance to return. Solving Eq.(8), we find

\ba
 N=N_{0} exp(-\tilde{\lambda} t)
\ea
where $\tilde{\lambda}=(1-\sigma)^{n} \lambda $, alternatively
$\hbar \tilde{\lambda}\equiv\tilde{\Gamma}=(1-\sigma)^{n}\Gamma$.
If $n=2200$, we have $ \Delta\Gamma/\Gamma=(\Gamma - \tilde{\Gamma})/ \Gamma =0.33 $,
which implies $\Delta \tau/\tau_{1/2}=0.492$. Let $t_{0}$ be the time
when the photon consumes during its round trip between the nucleus and
the plate, number of repeat of the process during the nuclear
half-life is $n=\tau_{1/2}/t_{0}=1542$. Since the program GEANT3 is
known to be efficient only for the gamma-ray energy larger
than $100KeV$, the value of $\sigma$ may be flexible for $81KeV$
gamma-ray.
For instance, $\sigma=0.00026$ renovates $n=1540$ to retain
$\Delta \Gamma /\Gamma =0.33$.
This analysis is anyway ad hoc.

\section{ Conclusion }
 In this paper, we report our discovery that the  decay of $^{133}Cs$
 in the first excited state was delayed when the nucleus was placed
 between two parallel flat plates.
 This phenomenon is based on the process that the population of
 the excited state of the nucleus is increased by reabsorption of the
 emitted photon into the same nucleus after being reflected by the
 plates.

\acknowledgements

 The author thanks Langouche and Milant in Leuven University for
 their help in the experiment.
 This work was supported by the Korean Ministry of
 Education(Grant no.98-015-D00061) and the Korean
 Science and Engineering Foundation(Grant no. 976-0200-002-2).

\bb{99}
\bi{1} Il-T. Cheon, J. Phys. Soc. Japan, {\bf 60}, 833(1991).
\bi{2}Il-T. Cheon, Hyperfine Interaction, {\bf 78}, 231(1993).
\bi{3} Il-T, Cheon, J. Phys. Soc. Japan, {\bf 63}, 47(1994).
\bi{4}Il-T. Cheon, G. Langouche and K. Milant, J. Phys. Soc. Japan,  {\bf 63}, 2453(1994).
\bi{5}Il-T. Cheon, Z. Phys. $ \bf D39$, 3(1997).
\bi{6}R. G. Hulet, E. S. Hlfer and D. kleppner
Phys. Rev. Lett. $ \bf 55$, 2137(1985).
\bi{7}G. Gabrielse and H. Dehmelt, Phys. Rev. Lett. {\bf 55}, 67(1985).
\bi{8}Jon J. Spijkerman, "An Introduction to M\"ossbauer
spectroscopy", ed. L. May, Plenum Press, 1971.Chapt.2
\bi{9}H. Pattny, private communication.
\bi{10} "Table of Isotopes, 8th Ed." Ed. R. B. Firestone and  V. S. Shirley,
Jhon Wiley \& Son, INC, New York, 1996.
\bi{11}J.J. Sakurai, "Advanced Quantum Mechanics", Chapt.2, Addison-Wesley Pub. Co. Inc.1980.
\eb

\newpage

Table 1. Changes of Decay Constant \\
\fbox{
$^{99in}_{43}Tc$ Technetium internal conversion
$\tau_{1/2}=6.007h$, 2.2KeV E3 transition }\\
\fbox{
Chemical
}
\begin{itemize}
\item{ 1953 K.Bainbridge et al.\\
$\frac{\lambda(KTcO_{4})-\lambda(Tc_{2}S_{7})}{\lambda(Tc_{2}S_{7})}=(2.70\pm0.10)\times10^{-3}$
}
\item{ 1980 H.Mazaki et al.\\
${[\lambda(TcO_{4})-\lambda(Tc_{2}S_{7})]}/{\lambda(Tc_{2}S_{7})}=(3.18\pm0.7)\times10^{-3}$\\
${[\lambda(TcS_{7})-\lambda(Tc_{2}S_{7})]}/{\lambda(Tc_{2}S_{7})}=(5.6\pm0.7)\times10^{-4}$
}
\item{ 1999 A.Odahara, T.Tsutsumi, Y.Gono, Y.Isozumi, R.Katana, T.Kikegawa, T.Suda, T.Kajino\\
${[\lambda(compound)-\lambda(metal)]}/{\lambda(metal)}=$in  progress
}
\end{itemize}
\fbox{
High pressure 10GPa
}
\begin{itemize}
\item{ 1952 K.Bainbridge et al.\\
${[\lambda(10GPa)-\lambda(0 Pa)]}/{\lambda(0 Pa)}=(2.3\pm0.5)\times10^{-4}$
}
\item{ 1972 H.Mazaki et al.\\
${[\lambda(10GPa)-\lambda(0 Pa)]}/{\lambda(0 Pa)}=(4.6\pm2.3)\times10^{-4}$
}
\end{itemize}
\fbox{
Low temperature
}
\begin{itemize}
\item{ 1958 D.Byers et al.\\
${[\lambda(4.2 K)-\lambda(293 K)]}/{\lambda(293 K)}=(1.3\pm0.4)\times10^{-4}$
}
\end{itemize}
\fbox{
External electric field gradient $\sim2\times10^{4}V/cm$
}
\begin{itemize}
\item{ 1970 H.Leunberger et al.\\
$\frac{\Delta\lambda}{\lambda}\approx 10^{-4}$
}
\end{itemize}
\fbox{
Phase transition : ferroelectric $\rightarrow$ para electric
}
\begin{itemize}
\item{ 1972 M. Nishi et al.\\
$\frac{\Delta\lambda}{\lambda}\approx (2.6\pm0.4)\times10^{-3}$
}
\end{itemize}
\end{document}